\documentclass[runningheads]{llncs}
\usepackage{amsmath}
\usepackage[T1]{fontenc}
\usepackage{graphicx}
\usepackage{amssymb}
\usepackage{comment}
\usepackage{mathtools}
\usepackage{multirow}
\usepackage{booktabs}
\usepackage{hyperref}
\footnotesize
\tiny
\begin{document}
\title{Precision ICU Resource Planning}
\subtitle{A Multimodal Model for Brain Surgery Outcomes}
\titlerunning{Precision ICU Resource Planning}
%
\author{Maximilian Fischer *\inst{1,2,3,4},
    Florian~M. Hauptmann *\inst{1},
    Robin Peretzke *\inst{1, 2},
    Paul Naser \inst{5,6}, 
    Peter Neher \inst{1,7},
    Jan-Oliver Neumann+\inst{5},
    Klaus Maier-Hein+\inst{1,3,7} 
}
\authorrunning{Fischer, Hauptmann, Peretzke et al.}

%
\institute{ 
Medical Image Computing Group, Deutsches Krebsforschungszentrum (DKFZ), Heidelberg\and
Medical Faculty Heidelberg, Heidelberg University, Heidelberg\and
German Cancer Consortium (DKTK), partner site Heidelberg\and
Research Campus M\textsuperscript{2}OLIE, Mannheim, Germany\and
Department of Neurosurgery, Medical Faculty Heidelberg, Heidelberg University, Germany\and
AI Health Innovation Cluster, German Cancer Research Center (DKFZ), Heidelberg, Germany \and
Pattern Analysis and Learning Group, Heidelberg University, Heidelberg
\email{robin.peretzke@dkfz-heidelberg.de}
}
\maketitle              
\begin{abstract}
Although advances in brain surgery techniques have led to fewer postoperative complications requiring Intensive Care Unit (ICU) monitoring, the routine transfer of patients to the ICU remains the clinical standard, despite its high cost. Predictive Gradient Boosted Trees based on clinical data have attempted to optimize ICU admission by identifying key risk factors pre-operatively; however, these approaches overlook valuable imaging data that could enhance prediction accuracy. In this work, we show that multimodal approaches that combine clinical data with imaging data outperform the current clinical data only baseline from 0.29 [F1] to 0.30 [F1], when only pre-operative clinical data is used and from 0.37 [F1] to 0.41 [F1], for pre- and post-operative data. This study demonstrates that effective ICU admission prediction benefits from multimodal data fusion, especially in contexts of severe class imbalance.
\end{abstract}
\newcommand{\asteriskfootnote}{\textsuperscript{*}}
\newcommand{\daggerfootnote}{\textsuperscript{+}}
\asteriskfootnote Equal contribution. Authors are permitted to list their name first in their CVs.
\daggerfootnote Equal supervision.
\section{Introduction}
Advanced neurosurgical techniques have significantly reduced the incidence of postoperative complications that require intensive care unit (ICU) admission, even among elderly patients \cite{3645-04}. Despite these advancements, many medical institutions still routinely transfer patients to the ICU following brain surgeries for the highest medical attention, even though ICU care is much more expensive \cite{3645-05}. Recent studies have challenged this practice in postoperative care, prompting a reevaluation of its necessity \cite{3645-01,3645-02}. 

A challenge in the field is the small number of patients who truly require ICU transfer, leading to few true positive cases in cohorts. Consequently, the class that demands the highest predictive accuracy is most impacted by the limited data available. The approach presented in \cite{3645-02} addresses this issue with a Gradient Boosted Tree (GBT) trained on clinical data and demonstrates better prognostic pattern recognition than traditional uni- or multivariate logistic regression. However, a limitation of previous approaches is its reliance solely on clinical data, overlooking valuable imaging information.

In this work, we enhance this predictive model by incorporating T1 MRI imaging data alongside clinical data, aiming to improve accuracy in ICU admission predictions, which has already been proven effective in other applications, such as Alzheimer’s disease prediction \cite{3645-03}. Furthermore, we especially focus on the evaluation of the model performances for true positive ICU subjects, which is of utmost importance.

Our study involves different architectures and approaches to fuse tabular information with imaging data. As baseline, we consider the same GBT approach as \cite{3645-02} and we use autoencoders (AE) to incorporate imaging data to generate latents from the input images.

In our work, we show that plane fusion of tabular and imaging data does not automatically lead to improved classification performances, even for deeper ResNet models than GBTs. For improved fusion of multimodal data for ICU admission prediction, dynamic fusion strategies, such as \cite{3645-03} are required. Combined with a brain foundation model as an image feature extractor, we observe promising results compared to existing GBT approaches and outperformed ResNet-based approaches.


To the best of our knowledge, this is the first approach to integrate multimodal data for predicting ICU admission in postoperative neurosurgical care.

\section{Dataset}
For our experiments, we used the same cohort presented in \cite{3645-02}. All subjects underwent brain surgery at the University Hospital Heidelberg, with various clinical parameters available, including demographic information and pre- and post-operative clinical data. Subject labels were assigned based on post-operative information, depending on whether one or more of nine events requiring ICU admission occurred post-surgery: CPR, re-intubation, return to the operation room, mechanical ventilation, vasopressor use, impaired consciousness, intracranial hypertension, swallowing disorders, or death. An overview of the available clinical information is provided in \cite{3645-02}.

\begin{figure}[b]
    \centering
    \includegraphics[width=\linewidth]{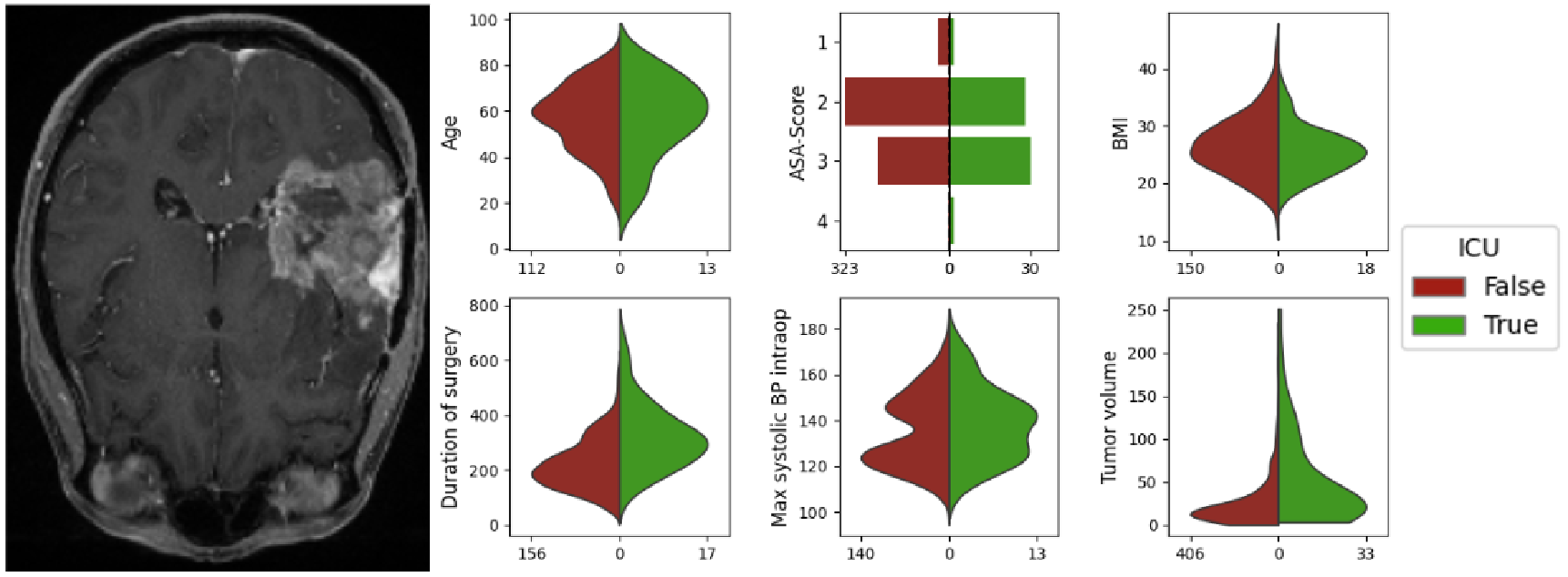}
    \caption{Left: Brain slice of a random subject of the cohort. Right: Overview of six tabular feature distributions.}
    \label{3645-fig-01}
\end{figure}

In contrast to \cite{3645-02}, we additionally incorporated imaging data, aiming to improve predictive accuracy. Since T1 scans were not available for all subjects that were used in \cite{3645-02}, our cohort was reduced to 611 subjects. For the remaining subjects, we created a segmentation mask of the tumor by using a pretrained nnUNet \cite{3645-06}. Within the dataset, a significant class imbalance was observed, with 552 ICU-negative and 59 ICU-positive labels, offering insights into the extent of potentially unnecessary ICU resource allocation. Fig. \ref{3645-fig-01} displays a random subject with a tumor, alongside example distributions from the tabular data of the cohort both before and after surgery.

\section{Materials and methods}
In this work, we employed multiple architectures to improve prediction accuracy. The used models can be distinguished into feature extraction and classification models.

\emph{Feature Extraction Models}: Feature extraction is a key method for aligning or reducing data dimensionality across domains. Here, we use AEs to derive compact representations from the image domain. These encoder-decoder models are pretrained on an image reconstruction task, compressing high-dimensional data into a lower-dimensional latent space that retains essential features while reducing overall dimensionality.

\emph{XGBoost}: Extreme Gradient Boosting is a decision-tree-based machine learning technique that efficiently processes 2D structured or tabular data. In \cite{3645-02}, this approach was applied to predict ICU admissions using clinical data. In our work, we extend this method by incorporating imaging based latents alongside the clinical data.

\emph{ResNet}: In contrast to XGBoost models, Residual Neural Networks are deep learning (dl) architectures with residual connections to facilitate the training of very deep networks. These connections enable efficient capturing of complex patterns. For our experiments, we use this model architecture as a more advanced approach than XGBoost.

\emph{Multimodal DAFT Model}: Along with introducing more advanced dl architectures, we hypothesize, that plane combinations of clinical and imaging data might not be sufficient for improved ICU-admission prediction. Thus we considered the Dynamic Affine Feature Map Transform (DAFT \cite{3645-03}), which enables the dynamic scaling and shifting of feature maps in a dl model, depending on the clinical data of a subject. This improved fusion of 3D imaging data with tabular clinical data has shown great results in predicting the outcome of Alzheimer's disease.

\section{Experiments}

This section provides an overview of the conducted experiments, divided into two parts: (i) feature extraction from T1-scans and (ii) the ultimate classification task. 
Each feature extractor was trained using five-fold cross-validation, generating latents only for the data in its respective validation fold. For the classification experiments, data splits were further stratified by label and tumor volume, and for the classification training suitable sampling strategies, such as undersampling, were applied to address the severe class imbalance in the data. A threshold of 0.5 was used to assign labels based on predicted probabilities. Evaluation metrics included F1-score and ROC-AUC, both of which are well-suited for capturing the effects of class imbalance.

\emph{Feature extraction}: For feature extraction, we implemented two approaches. First, we generated latent representations using our own pretrained 2D AE. Second, we leveraged a publicly available large-scale pretrained 3D model \cite{3645-07} (SSL). Both models were trained on image reconstruction tasks. The 2D AE was trained on our dataset using the ADAM optimizer with a learning rate of $1\text{e}^{-2}$, a batch size of 16, over 50 epochs. For the downstream task, we extracted latent representations from one axial slice of the T1 scan per subject showing the largest tumor diameter based on the segmentation mask, resulting in dimensions of (128 × 16 × 16).

For the 3D approach, we used the pretrained model from \cite{3645-07}, where the authors trained a masked AE on 44,000 multicenter images on different MRI modalities. This 3D model was fine-tuned during five fold CV on our data with a learning rate of $1\text{e}^{-5}$ over 15 epochs. For the classification task, we extracted latents from a 3D region of interest (ROI) centered on the tumor, defined by each subject's tumor segmentation mask. This ROI, sized at 160 mm³ with dimensions (320×5×5×5), was used to capture relevant latent features based on the T1 scan around the tumor. For clarity, we refer to the latents from the 2D AE as 2D latents and from the 3D AE as 3D latents.

\emph{XGBoost experiments}: For the GBT model, we conducted experiments using three different input scenarios: (i) solely on tabular data, (ii) solely on latents from the 2D images, and (iii) the concatenated tabular and flattened latent data from the 2D image. For all three scenarios, we used the same hyperparameters as \cite{3645-03}.

\emph{ResNet experiments}: The ResNet18-Classifier\footnote{\url{https://pytorch.org/vision/main/models}} experiments followed a similar structure, utilizing the same three input scenarios: tabular data alone, latent data alone, and a combination of both. For all scenarios, the data was shaped into a 2D array, using zero-padding. For all ResNet experiments, we trained 500 epochs using the ADAM optimizer with a learning rate of $1\text e^{-2}$ and a batch size of 32. 

\emph{DAFT experiments}: Since the DAFT model is specifically designed for multimodal purposes, we limited these experiments to combining clinical and imaging data. In our DAFT experiments, we integrated clinical tabular data with three types of latents: the 2D latents, where we stacked the same latent 16 times to meet DAFT’s dimensionality requirements; the latents generated by DAFT’s integrated 3D encoder; and 3D SSL latents. For all DAFT experiments, we used the ADAM optimizer with a learning rate of $1\text e^{-3}$ and a batch size of 16, training each DAFT model until convergence. The 2D model converged after 100 epochs, while the 3D model converged after 25 epochs, as the batchgenerators \cite{3645-08}framework was applied in the 3D scenario.

\section{Results}
\begin{figure}[b]
    \centering
    \includegraphics[width=\linewidth]{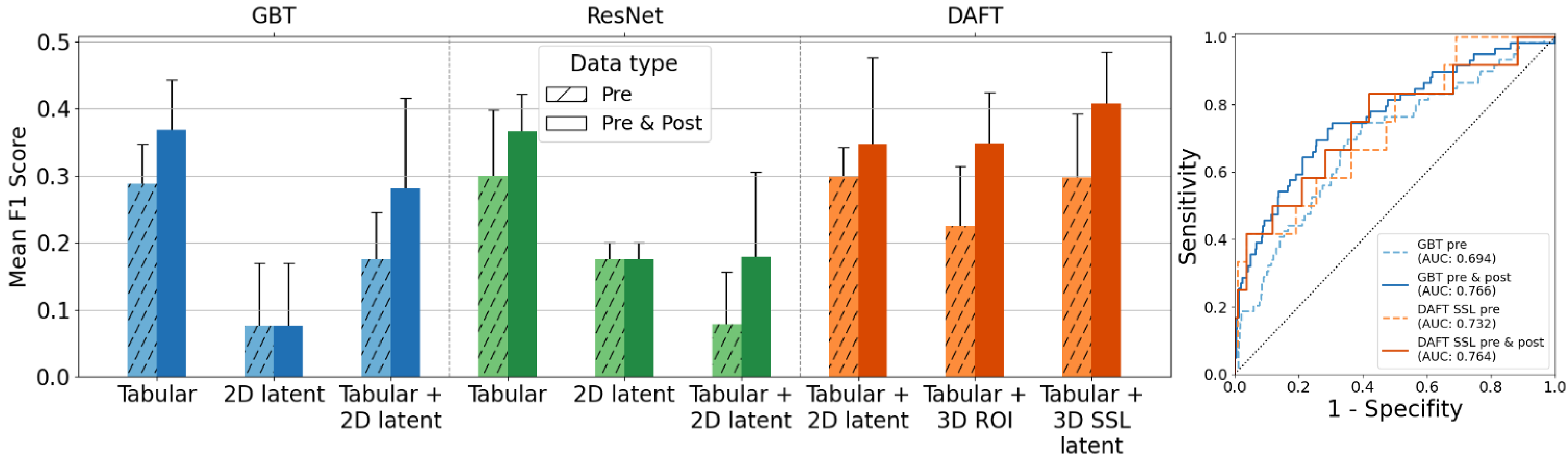}
    \caption{Left: F1-scores of different scenarios. The bar in the lighter color and dashed lines present the performance when only pre-operative data is used. Right: ROC-AUC curve of the GBT-baseline and our best performing method for only pre- and for pre- and post-operative data.}
    \label{3645-fig-02}
\end{figure}

As shown in Fig. \ref{3645-fig-02}, models incorporating both pre- and postoperative data demonstrated a consistent trend towards better performance compared to those using only pre-operative data across all pure tabular data scenarios. For instance, the GBT model, when using only tabular data, achieved an F1-score of 0.37 and a ROC-AUC of 0.77. However, when the GBT model relied exclusively on 2D latent features for classification, its performance dropped, with an F1-score of 0.08 and a ROC-AUC of 0.5. Combining tabular data with 2D latent features also did not improve the model’s performance over the tabular-only setup, yielding a lower F1-score of 0.28 and a ROC-AUC of 0.71.

With only tabular data it was similar for the ResNet: a F1-score of 0.37 and a lower ROC-AUC of 0.75. With 2D latent features alone, however, the ResNet classifier’s performance declined, reaching an F1-score of 0.18 and a ROC-AUC of 0.53. Combining tabular data with 2D latents led to further reductions, with an F1-score of 0.18 and a ROC-AUC of 0.66, performing even worse than the GBT under the same conditions.

In contrast, the DAFT model demonstrated the most potential for integrating tabular and imaging data. Using only 2D latents, DAFT achieved an F1-score of 0.35 and a ROC-AUC of 0.74, outperforming prior attempts at combining data types, though still slightly trailing the GBT trained solely on tabular data. When using a complete 3D ROI as input, the DAFT model maintained an F1-score of 0.35 but improved its ROC-AUC to 0.77. Finally, employing 3D latents from the pretrained SSL model yielded the best performance, achieving an F1-score of 0.41 and a ROC-AUC of 0.76.

\section{Discussion}
In our experiments, we progressively applied increasingly complex dl approaches to the problem of ICU admission prediction. Initially, we applied a GBT model trained on clinical data which, despite its outdated architecture, remains the most performant method currently published for this problem. In addition to clinical data, we have gradually integrated imaging data into the cohort, starting with 2D slices and extending to 3D volumes. To push the boundaries further, we applied a more contemporary architecture, ResNet, to leverage a deeper model with more parameters to extract richer correlations between the imaging and tabular data. However, without specific fusion techniques designed to effectively combine imaging and tabular data, this approach was limited due to the dimensionality differences between the two data domains.

The most promising approach, as revealed by our experiments, is the DAFT architecture. DAFT, with its specialized DAFT block, is specifically designed to fuse high-dimensional image data with low-dimensional tabular data (in our case, clinical data). Like other studies, our work was limited by the sparse availability of true positive ICU subjects. This is the main reason for generally low F1 scores across all methods. However, we were able to further enhance the performance by using a brain foundation model as feature extractor (see DAFT 3D ROI vs. 3D SSL latent). 

Another challenge for our work and other multimodal approaches, was the fusion of different domains. In our case, the tabular clinical data was tremendously smaller than the imaging data, even after feature extraction. For problems where prognostic markers for classification are present in smaller tabular data (see GBT on tabular data), it is important that these features are not lost in noise after being combined with image information. To solve this problem, we have combined two approaches: (i) the DAFT model which is capable of maintaining attention on low dimensional tabular data (ii) reducing noise in the image latents by using a foundation model as feature extractor. 

Our work was task-driven, aiming towards an improved ICU-admission scheme, especially for those patients that require ICU treatment. We have thus decided against exhaustive ablations, such as general 2D vs. 3D scenarios in our model development. In the future, we are planning to consider further experiments with 3D latents and ResNet models, other fusion strategies, and introducing more modalities such as different acquisition sequences to further improve the performance. Additionally, it is crucial to consider weighting strategies that prioritize accurately identifying true ICU patients. While misclassifying non-ICU patients as needing ICU admission may result in unnecessary resource allocation, the reverse—failing to identify patients who truly require ICU care—poses a serious risk to patient safety and could lead to fatal outcomes.

In conclusion, our experiments show that ICU-admission prediction is not a task that can be solved solely based on a single modality. However, combining different modalities can help to improve the classification performance. Moreover, large pretrained models as feature extractors are beneficial especially for heavy class imbalanced cohorts. Insights from this work can streamline future experiments to achieve clinical relevant results.

\newpage

\bibliographystyle{splncs04}
\bibliography{BibFile}
\end{document}